\magnification1200

\def\m{\mu}        

\def\Winf{$W_{\infty}\  $}
\def\winf{$w_{\infty}\  $}
%%%%%%%%%%%%%%%%%%%%%%%%%%%%%%%%%%%%%%%%%%%%%%%%%%%%%%%%%%%%%%%%%%%%%%%%%
\centerline{\bf REMINISCENCES}
\vskip0.4in

I was an undergraduate student at Kanazawa University, which had been
recently established as part of the post-war educational reforms.
Many of the professors had moved from the old imperial universities and
still followed the old curricula.
A few among them formed a research group for theoretical particle
physics. 
Since there were no students senior to us nor graduate students,
a few of us were made welcome in their reading club.
There, for the first time I was exposed to research in
theoretical particle physics.

Since Kanazawa university was an undergraduate college at that time,
I went to Nagoya university for my graduate studies.  Before going to
Nagoya I was already engaged in some calculations that I had been asked
to carry out by Oneda, my mentor at Kanazawa. 
These were $\Lambda - \beta$ decay calculations, and the results
were published in a joint paper by Iwata-Okonogi-Ogawa-Sakita-Oneda, a
paper that dealt with the universality of weak interactions and
in a sense, adumbrated the universal V-A interaction. 
The architects of the paper were 
Shuzo Ogawa and Sadao Oneda, from whom I learned the
phenomenology of strange particles (then called V-particles) and
the weak interactions.
This was about the time that the strangeness theory was
put forward by
Nakano-Nishijima and by Gell-Mann.

In Nagoya at that time, each graduate student belonged to a research
group. I belonged to Sakata's group. 
I stayed there for two years, and received a Master's degree
in 1956.  As I look back now, this was one of the
most fruitful periods for Sakata's group. The Sakata model was
proposed in my second year, although I was not a part of this
activity. By then I had become more and more interested in the collective
model of nuclei and the work in Nagoya by Marumori and others. My
master's thesis was on collective motions.

In 1956 I went to the University of Rochester.
When Robert E. Marshak, then
the chairman of the physics department at Rochester, visited Japan to
attend an
international conference in 1953, he expressed his interest in having a
number of Japanese graduate students join his group at Rochester. To this end,
he requested Yukawa and Tomonaga to select some students. Fourteen
students were selected between 1953 and 1959, I being one of them. I
received a research assistantship and a Fulbright travel grant.

In Rochester, I had to take regular courses during the first year. 
Among these courses, I had to take an experimental course; ``Modern Physics
Laboratory."  Before coming to Rochester I had been entertaining the
possibility of going over to experimental physics, 
as Koshiba and Yamanouchi
had done. After an unsuccessful X-ray experiment in that course,
however, I gave up that dream.

In the spring of 1957 there was a Rochester Conference which graduate students
were allowed to attend. A highlight of the conference was
Lee-Yang's work on parity non-conservation in weak interactions.
It was very exciting to see all these noted
people whom I knew only by name. There were many activities in
Rochester going on then too, such as the V-A theory of Marshak and
Sudarshan and the Marshak-Signel nuclear potential, to name a few.  
However, I, being timid and merely a new graduate student to boot, could not
participate in them.
It was very frustrating.

While I was at Rochester, I more or less followed the topic that was
current -- dispersion relations and symmetries (global symmetry,
pre-SU(3)). Dispersion relation was the subject that I had never
studied in Japan. While I was studying its techniques,
I tried to apply it to
various problems.
The first attempt was on the $K_{\m 2}$ decay where I mimicked the 
Goldberger-Treiman calculation of the $\pi $ meson decay. 
This calculation led to the conclusion
that the strangeness changing
current is much weaker than the strangness conserving current,
an indication of Cabibbo mixing.
Around this time, Marshak's group was working on
nuclear forces, computation of nucleon-nucleon phase shifts, the
photo-disintegration of deuteron, etc.  
J. J. de Swart, a student of Marshak,
was working on the photo-disintegration of the deuteron and
I was attracted to the subject. I
discussed with Susumu Okubo, a research associate then,
regarding the possibility of using dispersion relations in this problem
-- and eventually it became a part of my thesis at Rochester. 
My adviser was
Charles J. Goebel, then a young assistant professor, with whom I
finished my Ph.D. in 1959. He gave me complete freedom in physics
research and provided appropriate advice whenever 
necessary. During my three years at
Rochester, I received much encouragement
from my fellow Japanese graduate students,
and I also learned a great deal of physics
from them, especially particle theory
from Susumu Okubo and particle experiments from Taiji Yamanouchi.
%%%%%%%%%%%%%%%%%%%%%%%%%%%%%%%%%%%%%%%%%%%%
\vskip0.4in

I took a postdoctral job at the University of Wisconsin and moved to 
Madison in the summer of 1959. 
While I was in Rochester I did not work on weak interaction
phenomenology, but I maintained an interest in that subject. In Madison
I resumed research on
strange particle decays and collaborated with Oneda and Pati of Maryland
through
correspondences. This work is a precursor to the Penguin mechanism of the
$|\Delta I|=
{1\over 2}$ enhancement in non-leptonic weak interaction.

My boss at Wisconsin was Robert G. Sachs, who was keenly interested in
high energy experiments and had helped to build up a strong experimental group.
With his encouragement, I developed a friendship
with the experimental group, in particular 
with M. L. Good and W. D. Walker.
Through the next several years I closely observed the excitement they
felt in the
discovery of hadronic resonances. At about this time the Fry-Camerini
group errorneously claimed the discovery of $\Delta S =-\Delta Q$
events in neutral K meson dacays. Sachs had strongly supported 
the claim and had created such an atomosphere 
that for the next several years the theorists in his group
could not discuss models of
weak interactions without
$\Delta S =-\Delta Q$. Most of the models, based of the constituents
of hadrons such as the Sakata model or the quark model, became
extremely ugly. I spent fair amount of time and energy on the model
building of weak interactions only to be considerably frustrated. This was the
last time I was engaged in weak interaction physics. 

In Madison the Summer Institute was regularly organized by Sachs.
Attending the 1961 Summer Institute, I keenly felt my lack of knowledge
in group theory. I only knew angular momentum at the time.
In the summer of 1962 we did not have the Summer Institute, but
Jan Tarski, then a postdoc at the Institute for Advanced Study (IAS) in
Princeton,
visited Madison and he and I shared an office.
Since I knew that he had given a seminar on group theory in the previous year's
Summer Institute, I thought it was a good opportunity to learn group theory
from him.
Although I wanted to know the representations of SU(3), I asked him about
SU(4) instead, because I was shy about revealing my intention and moreover
thought that if I understood "4" I would understand "3". He then started
explaining it at great
speed by drawing a lot of dots on the blackbord. 
But, when he found me completely
foxed, he looked at me pityingly and said that the
representations of SU(4) had all been worked out by Winger.
This is Wigner's famous paper on the supermultiplets of nuclei.
Immediately I went to the library.
Although I did not quite understand the group theory part of the paper,
I understood very well the intent of the paper, in particular
the introduction, where
Wigner explains
why and how group theory is applied to this problem.

My teaching career began in 1963. That year I
taught a course called ``Special Topices in Theoretical Physics"; the weak
interactions in the fall semester and the strong interactions in the spring
semester.
This was
the year of SU(3); the Cabibbo theory of weak interactions
and the quark model.
I reviewed in detail
the status of elementary particle research at that time
including some
of my own contributions.
In the last lecture of the course I discussed the non-relativistic SU(6)
theory, a supermultiplet theory of hadrons based on the non-relativistic
quark model, which I
had developed during the second term (Spring 1964).
I was well aware of the limitation of the theory, especially the difficulty
associated with the relativistic generalization.
I worked on this problem all the time during the summer of 1964.
I presented my version of SU(6) in the last seminar of that year's Summer 
Institute.
Louis Michel and Eugene Wigner were there in the audience. After the seminar
Wigner came back to the seminar room and informed me of
the news he had received from G\"ursey and Radicati on their independent work.
Next day Wigner invited me to his office, and grilled me with several questions
before convincing himself that I had used the same SU(6) group as had
G\"ursey and Radicati, and then he opened a briefcase
to show me their paper. By then I knew the content
of the paper because I had called Feza G\"ursey
the previous day and had learned it from him. There was an important difference,
however, between the two. The difference was that
I had chosen the anti-symmetric representation "20" of SU(6) for baryons based on the
naive quark model,
while they correctly
chose the symmetric representation "56".
During the summer I had been so preoccupied with the relativistic problem
of SU(6)
that I had neglected to examine the SU(3) contents of some of the possible
representations of SU(6). When I expressed my concern about the
relativization, Wigner politely refused the discussion and
suggested that I discuss the matter with Louis Michel.

A few days after the seminar, I moved to Argonne National Laboratory, where
Sachs became the associate director of the laboratory
in charge of high energy physics research
and I joined its theory group. The first thing I did in Argonne was a quick
calculation of the
nucleon's magnetic moment using the representation "56". When I found the ratio
$-{3/ 2}$
for the magnetic moment of the proton and the neutron, I became
fully convinced of the correctness of the
"56", and thought that we might have to abandon the naive quark model.
As is well-known, the resolution of this dilemma later led to
the introduction of the color degree of freedom, and eventually to QCD.

In Argonne, before my family and I found a house in a nearby town, we stayed
for one month at the visitors' housing facility. Harry Lipkin, who showed
a strong interest in SU(6), was also there with his family.
A few days later Michel and his family moved
from Madison and stayed in the facility for a month or so before going back
to France. I started to discuss with Michel about the feasibility of 
relativistic
SU(6). But we had communication problems, because his group theory was much too
sophisticated for me. Nevertheless, we managed to finish a work,
which dealt with a discussion of, together with some negative results in
the relativistic extension of the SU(6). A joint paper, in actuality
written by Michel entirely in his own style, was drafted.
To tell the truth, he had agreed
to write a more elementary paper with me by using Lie algebra. But,
we have never finished it mainly because
he was satisfied with his version and I became extremely busy working on
the next project with Kameshwar C. Wali.

This was the time when Argonne played the role of a center of
high energy physics activities in the Midwest. Many physicists from the
nearby universities
gathered at the weekly seminars.
Yoichiro Nambu and Jun J. Sakurai showed up quite frequently. I had known Nambu
for some time since he was a frequent visitor to the Summer Institute at
Madison.
In discussions with him I myself became convinced that an attempt
at a phenomenological but
relativistic formulation of SU(6) would be worth a try, inspite of the
negative results we had had regarding SU(6) being an essential
theory.
K. C. Wali and I started working on this intensively in the late fall of 1964,
and we finished the work just before
the 1965 Coral Gables Conference in January, to which I had been invited.
The first speaker of the conference was Abdus Salam, who spoke on
the relativistic $\tilde{U}(12)$ theory of Salam-Delbourgo-Strathdee.
I was shocked by the talk since their work was identical to ours, even to the
notations.
During the talk W. D. McGlinn, who knew our work and was sitting in front of me,
turned around several times and kept gesturing to me to speak up.
After the talk I spoke up and then handed a hand-written copy of our paper to
Salam.  At a party that evening Salam returned the paper back to me saying
that both works were the same. And then he invited me to visit the
International Center for Theoretical Physics (ICTP) where he was the director.
In the following week at the APS meeting in New York,
Pais presented a work with Beg on their version of 
the relativistic SU(6), 
which subsequently was reported in the New York Times.
A few weeks later, I saw Salam's picture
in an English newspaper that had been posted on a
bulletin board in Argonne. 
I was not too happy about the fact that our contribution 
was not suitably acknowledged. Moreover, the publicity accorded 
to a work
that I regarded as merely phenomenological, made me quite uneasy.

In the same Coral Gables conference, Roger Dashen gave a talk on the
bootstrap program, in which he was describing 
the various vertices in terms of the matrix elements
of a few matrices.
After I came home I realized that Goebel's strong coupling theory
could be formulated algebraically by using matrices, 
and the next moment
I had obtained a Lie algebra of the symmetry group  
which serves as a spectrum
generating algebra for isobar states (hadrons). I immediately informed
Goebel, my advisor at Rochester, of this development. In the meanwhile, in 1961,
he had joined the faculty
at Wisconsin.
In Argonne at that time there were several group theorists: 
Morton Hammermesh,
William D. McGlinn, and the mathematician Robert Hermann.
When Michel was still there, he proposed having a series of 
tutorial lectures by Hermann. 
Based on these lectures Hermann later wrote his well-known
Benjamin book.
When I told Hermann about
my findings on the strong coupling group (as we named it), 
he showed a strong 
interest in it and suggested that I study a few
relevant mathematical subjects:
group contraction, induced representation, and the Peter-Weyl theorem.
This was a very valuable suggestion, for
Thomas Cook (my first student) and I spent the next one year or so working on
these problems. Meanwhile, we published a short paper (Cook-Goebel-Sakita)
on the strong coupling group together with a 
derivation of the representations for a few simple cases by
using the method of contractions.

Suddenly the SU(6) became fashonable and I started receiving many invitations.
In the summer of 1965 I was invited by Delhi University to give a series
of lectures in a summer school at Dalhouse, 
a hill station in the Himalayas.
Using this opportunity, I traveled around the globe: 
India, Japan and England
in that order to participate in summer schools and conferences.
This visit to Japan was the first since I had left nine years previously.
I was well received there and
I really felt the difference that the SU(6) had made.

%%%%%%%%%%%%%%%%%%%%%%%%%%%%%%%%%%%%%%%%%%%%%
\vskip0.4in

In the spring of 1967 I stayed at the ICTP in Trieste
for five months. Towards the end of the stay K. C. Wali and 
I traveled to
Israel, specifically the Weizmann Institute for ten days at the invitation
of H. J. Lipkin.
When we arrived in Israel we found that the atmosphere
was extremely tense and people busy preparing
for a war with the neighboring Arabic
countries. Although the touristy places were deserted,
we could manage to rent a car to visit many places including
Jerusalem, Haifa and Acre.
Since most of the young Israeli physicists had already been
drafted, the physicists working at the Institute were mainly
foreigners, among whom were H. Rubinstein, G. Veneziano and M. Virasoro.
They were working together on superconvergence relations, which was a
subject that I was also interested in at that time. In 
the discussions we had during this visit,
the dual resonance program must have come up,
since I remember that afterward in Trieste I started discussing
with others about the possibility of constructing
scattering amplitudes by summing only the s-channel resonance poles.
We left Israel as scheduled on June 4 and the very next day in Ankara, 
Turkey we heard of the outbreak of the Six Day War.

In 1968, Keiji Kikkawa and Miguel
Virasoro, with whom I had become acquainted in the previous trips
to Japan and to Israel, joined our group at Wisconsin as research associates.
By then I had returned to the University of Wisconsin to resume teaching,
which I missed at Argonne, and I was preparing a course,
``Advanced
Quantum Mechanics", which was essentially a one-year graduate course on
quantum field theory. 
In that summer
Virasoro showed up in Madison with a hand-written
paper by Veneziano and he explained to us in detail the activities of 
the Weizmann Institute.
At once Goebel and I got interested in the work and all of us
started thinking about generalizations.
In that fall after Virasoro succeeded in obtaining the five point Veneziano
formula, our
activity became intensified and within a few weeks Goebel and I had obtained
the $N$ point Veneziano formula. Then Kikkawa, 
Virasoro and I started to generalize
the formula further to include loops.

At this point we faced a dilemma. Namely, if one considered the Veneziano
formula
as a narrow resonance approximation of the true amplitude as was commonly
assumed at that time,
the construction of loop amplitudes
based on this approximate amplitude did not make sense.
After reviewing the logistics of quantum field
theory,
we arrived at the conclusion that the construction of the loop amplitudes
did make sense if we
considered the Veneziano amplitude
as a Born term of an unknown amplitude for which we had
an expansion similar to the standard Feynman-Dyson
expansion in perturbative field theory.
With this philosophy in mind we decided to construct 
a new dynamical theory of strong interactions.
First we defined the
local duality transformation as the crossing transformation
at any four point sub-diagram of a Feynman
diagram, and invented
a Feynman-like diagram compatible with duality as a diagram which
contained all the Feynman diagrams related to each other by 
the local duality transformations.
Then we simply wrote down a prescription for the
scattering formula corresponding to each of these Feynman-like diagrams.
In practice, we used diagrams which were dual to the Feynman diagrams.
A three point vertex of a Feynman diagram corresponds to a 
triangle in the dual diagram. 
An $N$ point Feynman tree diagram corresponds
to a specific triangulation of an $N$ polygon in the dual diagram.
A local duality transformation in the dual diagram is the transformation
of one triangulation of a quadrangle to another triangulation.
In terms of the dual diagram, therefore, an $N$ point Feynman-like tree diagram
corresponds to an $N$ polygon.
By studying these Feynman-like diagrams, it became clear to us that
a dual amplitude corresponded in a one-to-one fashion to a two-dimensional 
surface with
boundaries, and equivalently to a Harari-Rosner quark line
diagram, which, by the way, we had also invented independently.
In the second paper, we discussed
the general Feynman-like diagrams by using the classification of two
dimensional surfaces, and extended the prescription to non-planar diagrams.
This classification is the same as that of open
string amplitudes.

Kikkawa left Madison in the summer of 1969 for Tokyo, and Virasoro
and I left the following summer bound for Berkeley and France
respectively.
By then the operator formalism of the dual resonance amplitude had been
established by
S. Fubini, D. Gordon, and G. Veneziano, and independently by Y. Nambu,
who further proposed the string interpretation based on this work.
When I heard of the string interpretation from Nambu 
I felt it as natural as if I had known about it beforehand. 
I remember that I had experienced the same feeling
when I had first heard about the Sakata model from Sakata.

At Wisconsin Virasoro had used the operator formalism to analyze
the possibility of the negative metric ghost states consistently
decoupling from the physical states. He obtained a set of operators, which
could be used consistently as the operators of subsidiary conditions on
the physical states. These operators were later found to be the generators of
conformal transformations on a complex plane. 
This is the origin of the Virasoro algebra.
In discussing this problem with him, I realized that these operators
were compactly expressed in terms of a scalar field in a fictitious
1+1 space(finite)-time, and the Veneziano formula itself could
be expressed in terms of this scalar field operator. At about 
this time we received a hand-written paper by H. B. Nielsen:
``A physical model for the $n$-point Veneziano model."
Inspecting a few mathematical formulae in the paper, 
I came up with a functional integral representation of the Veneziano formula. 
There remained several important points to be clarified,
such as the M\"obius invariant property of the functional
integrand, the connection with the operator formalism, and the calculation
of non-planar amplitudes. At the end I, 
together with Virasoro and my student C. S. Hsue,
established the functional path-integral formulation of dual resonance
amplitudes, and with Virasoro, a physical
model of the dual resonance model based on the ``fishnet" diagram.

I stayed in France for one year before 
I moved to the City College of New York
in 1971.
To lessen the financial burden on the Institute, 
Michel had arranged a joint
invitation
by his Institute at Bures-sur-Yvette 
and the Bouchiat-Meyer group at Orsay, a group
of physicists that later moved to the \'Ecole Normale Sup\'erieure in Paris.
This arrangement turned out to be a very fortunate one for me, as
in Orsay I found several young physicists, who were interested in our
work. Moreover it was there that I succeeded in starting
a long and fruitful collaboration with Jean-Loup Gervais.
During this visit, I wrote three papers with Gervais:
on the functional integral, conformal field theory, 
and the super-conformal-symmetry,
all in conection with the dual resonance model.
   
In Wisconsin I had already started working on the
factorization of dual resonance amplitudes using the slicing and
sewing technique of functional integrals. I had drafted the preliminary results
into a paper and had
sent it to the Physical Review before I arrived in France. 
At Orsay, however, I withdrew the paper,  
as a result of discussions with Gervais,
when I convinced myself that a part of the paper was wrong.  
There were plenty of technical difficulties, on which
Gervais and I had to spend another half a year of hard work.
In this work we used formally and fully the conformal 
transformation properties
of functional integrals without seriously questioning their validity.
Sometime later we suspected the existence of an anomaly,
that would explain the critical
dimension of the model. I regret that we did not pursue it further.

When I received a paper on the new dual pion model
of Neveu and Schwarz in the spring of 1971, 
I noticed at once that the most important
ingredient in the model was the conformal invariance property.
One could discuss about the generalization of dual resonance amplitudes
in the very general terms
of conformally invariant field theories. 
So, Gervais and I got busy constructing
conformally invariant field theories. 
In this work, we discussed first a general theory of conformal
fields by defining the irreducible fields 
(now known as primary fields)
and the conformally invariant Lagrangian, and then we
established the functional-integral representation of
Neveu-Schwarz model by introducing a fermionic field in the model
in addition to the old bosonic field. 
After the work was completed I wrote a letter to Virasoro (in Berkeley then)
informing him of our work,
since I heard that he
had presented
a similar work at a conference in Israel. 
In the exchange of letters,
I learned the Ramond model from Virasoro and that
it also could be described
by the same Lagrangian simply by changing the boundary condition
on the fermionic field.

Gervais and I thought that in the functional-integral
representation the elimination of ghost states could be done by factoring
out the negative metric components of the fields by using conformal
transformations as was done in the standard gauge field theories. 
The necessary condition for this is, of
course, that the Lagrangian is
invariant under the conformal transformations.
Once we introduced a new field in the new model which generated new ghost
states, we had to find out a new set of gauge transformations under which the 
Lagrangian was invariant.
Neveu-Schwarz-Thorn had just published a paper in which they proposed
a set of operators to be used as the subsidiary gauge conditions
on the physical states of the dual pion model.
We tried to interpret these operators as the Fourier modes of 
the Noether current
associated with the new gauge transformations which involved
the new fermionic field, and
arrived at
the superconformal gauge 
transformations, under which the Lagrangian we had obtained previously 
was invariant.
I believe that these field transformations are the
first instances of supersymmetry transformations in
a local field theory.
The day after we had drafted this paper, I left France for New York.
In this work, we had to use anti-commuting c-numbers (Grassmann numbers)
and functional integration of fermionic variables. These, to us were new
concepts and we were initially reluctant to use them. Apparently, others
shared this reluctance and this work and the functional-integral work in
general, was not appreciated in our circle.
However, I received an impression 
that when
I presented the work later in December at the conference on 
functional integration at
the Lebedev Institute in Moscow, it, as well as
the use of anti-commuting c-numbers was well appreciated.

When R. E. Marshak became the president of the City College in 1970,
I, together with Keiji Kikkawa, accepted a position there.
I continued my research on dual resonance theory for a few more years,
after I had settled down in the City College.
There was a big difference, however, between before and after coming to
the City College.
Although several faculty members were already there before I came, 
I was expected
to play the role of the leader of the high energy theory group.
I felt that it was a great challenge to elevate the group into a quality
research group.
In a few years,
thanks to Marshak's personal connections,
we could gather a few talented graduate students into our group.
And also we could hire a new faculty member, Michio Kaku and postdocs,
such as
Yoichi Iwasaki. Moreover, I could invite
J.-L. Gervais for short visits on a few occasions.
I intentionally spent more time with students, and shared my
insights with them.

In the early spring of 1973, I was invited by Ziro Koba to visit the Niels Bohr
Institute in Copenhagen for two weeks to deliver a colloquium, and
more importantly to discuss the dual resonance string theory with his
group, in particular with Holger B. Nielsen and Paul Olesen. By this time at
the City College, Gervais and I had already formulated 
the ghost free Veneziano amplitudes by using the functional-integral 
representation of the Nambu-Goto string in the light-cone gauge.
This work later led to Mandelstam's
factorizable functional formulation of light-cone string theory,
and eventually to Kaku-Kikkawa's
light-cone string field theory.
Furthermore in our group at that time, 
the work of Iwasaki-Kikkawa was near
completion. This was
an attempt, which I persuaded them to carry out,
at a formulation of a light-cone string theory for the
Neveu-Schwarz model.
I reviewed these activities in Copenhagen. While I was in Copenhagen, David
Olive
called me up asking me to visit CERN on the way back home.
At the CERN seminar, I reviewed the Iwasaki-Kikkawa theory.
Later, I was told that this seminar and a conversation after the seminar 
had led
Wess and Zumino to start their seminal work on supersymmetric field theory.
I vividly remember the conversation with Zumino at the CERN coffee lounge.
When I said, ``If you allow
me to use anti-commuting c-numbers, 
Gervais and I have written down 
a transformation of a fermi field to a bose field in the Nuclear Physics paper",
he replied,
``It's OK to use anti-commuting c-numbers. Schwinger has frequently
used them."

%%%%%%%%%%%%%%%%%%%%%%%%%%%%%%%%%%%%%%%
\vskip0.4in

In the June of 1968 there was an international symposium at the ICTP
celebrating its new building at Miramare.
At the symposium I was introduced to Faddeev and
from him I learned the Faddeev-Popov trick.
Being fascinated by the method I tried to use it in
various problems, and gradually I convinced myself that the method could be
useful
for a much wider class of problems than simple gauge fixing.

My encounter with the strong coupling theory of Wentzel
goes back to my student days at Rochester. Since then I had been
observing the development of Goebel's S-matrix approach to the strong
coupling theory from up
close. As I have mentioned before,
I even contributed to it by formulating and extending Goebel's
theory in the form of an operator algebra including multi-partial
waves. Through this work, I became acquainted with Gregor Wentzel and
I was even introduced as his grandson at his retirement dinner party
in Chicago,
since Goebel was his student.
But, to tell the truth I had never seriously studied
his field theory of the strong coupling model. 
When I learned the Faddeev-Popov
trick, it occurred to me to develop a functional-integral formulation
of the strong coupling theory by using this trick.
Because of other work that had to be done meanwhile, I could not
even get started on this project
until I had found two students, Gustavo C. Branco and Pavao Senjanovic,
at the City College.
At the beginning I thought that the problem was rather
easy and one that was appropriate for
graduate students. It turned out, however, that we had to
overcome many obstacles; of which some were crucial
albeit most were technical.
I remember that
I had to read Tomonaga's strong coupling paper again very carefully.
At the end we succeeded in
the functional-integral formulation of the strong coupling theory.
There were two important general issues involved in this work,
namely, (a) the introduction
of collective coordinates in field theory, and (b) the semi-classical
expansion in field theory.
But, I suspect that at that time we did not fully recognize
the relationship between the collective coordinates and the zero modes, nor
that between the
strong coupling limit of the static models and the classical limit.

In the summer of 1974, I went to Europe to attend the
International Conference on High Energy Physics in London. Before the conference
I stayed in Orsay for a month. During this time, influenced by a seminar
given by Neveu on his work with
Dashen and Hasslacher, Gervais and I
decided to work on the semi-classical quantization
of classical solutions, in particular the Nielsen-Olesen vortex solution
of the Higgs model. Gervais studied our strong
coupling paper very carefully and brought his
new insights to bear upon it. In this summer we worked together
in Aspen and at Brookhaven successively for several weeks to finish
up this work on the quantized relativistic string
as a strong coupling limit of the Higgs model.
In this and in the subsequent work on
soliton quantization, we used the Faddeev-Popov
trick to extract the collective coordinates
out of bosonic field theories.
With this as the starting point, Gervais and I, together with Antal Jevicki
(then a student) had firmly established the collective coordinate method
as a method of semi-classical expansion in field theories, by the time of
the following year's workshop on
``Extended System in Field Theory"
held at the \'Ecole Normale Sup\'erieure.

In these works we performed point canonical transformations in
the functional-integral representation of
bosonic quantum field theories. Since in functional-integral representations the
operator ordering is not explicit, one often misses
a term which is
proportional to $\hbar$. Sure enough, we missed such a term in our work as
was pointed out by E. Tombolis.
Although Gervais and Jevicki have shown subsequently that
it is possible to incorporate operator ordering into
the functional-integral formalism, I realized that this was
a serious drawback in the actual application
of the functional-integral formalism.
It was then that I decided to use the Hamiltonian operator formalism
whenever a change of variables in quantum mechanics was involved.

Meanwhile, I had received a paper from Kikkawa, who had returned to Japan
the previous year; the Hosoya-Kikkawa paper on the
gauge theory
of collective coordinates.
The main idea of this paper was to construct for a given theory
an artificial gauge theory,
which involved the collective coordinates as gauge parameters, such that
if one fixed the gauge by setting the collective coordinates zero,
the theory reverted back to the original theory.
A natural question occurred to us: what would happen if one applied this
method to
a genuine gauge field theory?
It turned out that the most of the collective coordinates were absorbed into
the vector potentials by gauge transformations except for the collective
coordinates at the boundary of the system, which manifested themselves as 
surface variables. Subsequently we, Gervais-Sakita-Wadia, found that these
surface variables were indispensable for a gauge invariant quantum mechanical
description of the 't Hooft-Polyakov-Julia-Zee monopole-dyon solution. 
I encouraged Spenta Wadia (then a student) to investigate this problem further,
addressing non-Abelian gauge theories in general, by using
the Hamiltonian formalism.

At about this time Gervais and I
worked together fairly regularly.
According to my notes, we worked together in New York, Aspen,
and Paris for a total of twelve weeks in two years ('76-'77).
We developed the many-variable WKB method,
the $A_{0} =0 $ canonical formalism for non-Abelian gauge theories,
and together
with H. J. de Vega, a real time approach to instanton phenomena.
In our group at the City College around this time,
R. N. Mohapatra, who succeeded Kikkawa, was actively working with his students
on his left-right symmetric
model of weak interactions.
Michio Kaku was productive in conformal supergravity with
Townsend and van Nieuwenhuizen of Stony Brook.
The main theme of the research surrounding me was
the non-perturbative study of non-Abelian gauge field theories.
Tamiaki Yoneya ( then a postdoc ) and Spenta Wadia
were very active.
I recall one of their works 
that dealt with the role of surface variables in the vacuum structure of
Yang-Mills theory. In this paper, they explicitly transformed 
the Belavin-Polyakov-Schwarz-Tyupkin solution to the Coulomb gauge. 
They obtained a pendulum
equation for the transformation function, which led to infinitely many 
solutions. This infinite multiplicity is now known as the Gribov phenomenon,
but their paper predates that of Gribov by almost a year.
 
%%%%%%%%%%%%%%%%%%%%%%%%%%%%%%%%
\vskip0.4in

Large N QCD had been introduced by G. 't Hooft in 1974.
One of his motivations was to remedy the arbitrariness involved
in the "fishnet" diagrammatic approach to the string theory of strong
interactions.
This and the subsequent developements
influenced us into thinking about the large N expansion in field theory
in general.

In the winter of 1978-79, Wadia, then a postoc at the University of Chicago,
came back to New York and informed me of a work with Eguchi.
This discussion stimulated me to think about a gauge invariant calculation
of non-Abelian gauge theories. I tried to rewrite a non-Abelian gauge theory in
terms of equal-time Wilson loop variables. For this purpose I used the
Hamiltonian
canonical formalism and
the method of change of variables that I had learnt long ago in Nagoya.
To my surprise I obtained a field theory of interacting strings as a large N
limit of the SU(N) gauge theory.
In order to justify the procedure, I applied the same procedure to known
examples:
a collection of many identical free harmonic oscillators, and high density
bosonic plasma
oscillations. It worked correctly only if I made a similarity 
transformation such
that the resulting Hamiltonian became hermitian. I was mystified by this
until I spoke with Jevicki, then a postdoc at the IAS, who pointed
out 
at once that this transformation essentially took care of the contribution from the 
Jacobian of the change of variables.
So, I proposed to him that we write a joint paper on the general theory of the
collective field method, so named later, after polishing up all the
calculations.
When I was writing a first draft of the paper in the summer of 1979, 
Jevicki informed me of the work of Br\'ezin-Itzykson-Parisi-Zuber 
on the large N quantum matrix model, to which we could apply our method.
Indeed, it was not difficult to derive their
result by the collective field method.

As I have mentioned above, in Nagoya I had learned the method of change of 
variables
used in the collective
field theory. I was, and still am, curious as to whether
I had a prototype of the collective field
theory in my master's thesis or not.
Sometime later when I went back to Nagoya, I went
to the library to look for the thesis. However it was missing from the library.

In the fall of 1983 Gervais stayed in the City College for a few months
as a visiting Professor to fill in the gap created by the departure of
R. N. Mohapatra.
I do not remember why, but I was explaining to him the derivation of the
strong coupling
group and its representation by the method of
group contraction, namely the old work of Cook-Goebel-Sakita.
To my surprise, he was visibly excited.
He said `` This could just be the large N QCD."
It did not take us long to realize
the relation between the large N baryons and the strong coupling theory:
$\sqrt {N} \approx {G} $, once we learnt that
Witten had shown that in the large N QCD
the masses of baryons are of order $N$ while
the meson-baryon Yukawa couplings are of order $\sqrt N$.
We spent several more weeks to complete the work,
as we tried to establish its relation to the solitonic Skyrmion physics:
$\sqrt {N} \approx {G}\approx {1/g}. $
I remember, when we had finished the work,
that I was extremely satisfied with it, since it involved many
of my previous works; SU(6), strong coupling algebra, and soliton quantization,
which were seemingly unrelated until then.
%%%%%%%%%%%%%%%%%%%%%%%%%%%%%%%%%%%%%%%%%%%%%

\vskip0.4in

In the fall of 1980 I spent 4 months at the Yukawa Institute in Kyoto.
On the way to Japan I was in Europe for several weeks and
I came across the Parisi-Wu stochastic quantization paper
in CERN's preprint library.
Although I had been interested in statistical
mechanics, I had never seriously worked on the subject. When I came back to
New York the next spring I decided to spend some time in studying 
non-equilibrium
statistical mechanics and stochastic processes in particular. 
My source was a Japanese book
by R. Kubo and M. Toda entitled {\it Statistical Physics}, one of the Iwanami 
series, which I
had bought
in Japan. I translated one relevant chapter on stochastic processes
and distributed it among my students, Guha, Alfaro and Gozzi,
as they were all fascinated by
stochastic quantization. Over the next few years several papers on stochastic
quantization appeared from
our group: on large N reduction by Jorge Alfaro and me,
on supersymmetry and stochastic quantization by Ennio Gozzi,
on stochastic quantization of supersymmetric theories by Kenzo Ishikawa,
and
the calculation of the chiral anomaly by Rodanthy Tzani.

At about this time I taught on two occasions a special topics course on
Field Theory and
Statistical Mechanics, which included such topics as the derivation of the
Landau-Ginzburg equation in the BCS model and the Lee-Low-Pines
theory of polarons
in terms of Feynman's variational method. Based on the lecture notes
compiled by the students, I wrote a book called {\it Quantum Theory of
Many-Variable Systems and Fields}, World Scientific Lecture Notes in Physics
Vol.1, published in 1985.

In the spring of 1985 I met
Zhao-bin Su of the Institute of Theoretical Physics
in Beijing, who was a visitor to  our condensed matter
theory group at City College.
He introduced me to several topics in
condensed matter physics and frankly revealed the problems
he was facing.
One of them concerned the charge density wave transport phenomenon
in a one dimensional system of electrons in a crystal.
During the exchange of questions and answers, we gradually
realized the importance of chiral symmetry and the chiral anomaly
for this phenomenon.
This work, together with a later work with Kenichi Shizuya on the same subject,
drastically changed the level of my understanding of the chiral anomaly
and the physics of anomalies in general.
Another topic Su had introduced me to was the fractional quantum Hall effect.
I have spent a fair amount of time and energy on this subject
over the past ten years,
and have written several papers with him, but I
must confess that to this date I still do not understand the subject to my
satisfaction.

In the past ten years I have become more and more interested in
many-body problems,
which is the subject I was involved in when I was a student at Nagoya.
Interestingly, to me, the subject is rich enough to fill the gap between
condensed matter physics research and particle physics research.
When I learnt about the  \Winf algebra in a seminar on string theory,
I realized that
I had obtained the same algebra in my study of the fractional quantum Hall
effect.
I simply had not thought about the significance of the algebra in the
physics of the Hall effect.
In this respect
I am pleased with a series of works I have done over the past five years
with Dimitra Karabali, Satoshi Iso, and
Rashmi Ray, since all of these works illuminate
the significance of the \Winf ( or \winf ) algebra in the physics of
low dimensional fermionic systems.
%%%%%%%%%%%%%%%%%%%%%%%%%%%%%%%%%%%%%%%%%
\bigskip
\ \ \ \ \ \ \ \ \ \ \ \ \ \ \ \ \ \ \ \ \ \ \ \ \ \ 
\ \ \ \ \ \ \ \ \ \ \ \ \ \ \ \ \ \ \ \ \ \ \ \ \ \ 
\ \ \ \ \ \ \ \ \ \ \ \ \ \ \ \ \ \ \ \ \ \ \ \ \ \ January 1997
in New York

\ \ \ \ \ \ \ \ \ \ \ \ \ \ \ \ \ \ \ \ \ \ \ \ \ \ \ \ \ \ \ \ \ 
\ \ \ \ \ \ \ \ \ \ \ \ \ \ \ \ \ \ \ \ \ \ \ \ \ \ 
\ \ \ \ \ \ \ \ \ \ \ \ \ \ \ \ \ \ \ \ \ \ \ \ \ \
\ \ \ \ \ \ \ \ \ \ \ \ \ \  Bunji Sakita

\end